\documentclass[10pt, a4paper, conference, compsocconf]{IEEEtran}
\usepackage[english]{babel}
\usepackage[utf8]{inputenc}
\usepackage[T1]{fontenc}
\usepackage{amsmath}
\usepackage{amsfonts}
\usepackage{amssymb}
\usepackage{framed}
\usepackage{dsfont} 
\usepackage[standard]{ntheorem}
\usepackage{subfigure}
\usepackage{booktabs}
\usepackage{color}
\usepackage{calc}
\usepackage{stmaryrd}
\usepackage{alltt}
\usepackage{tabls}
\usepackage{textcomp}
\usepackage{times}
\usepackage{tabularx}
\usepackage{sublabel}
\usepackage{ctable}

\begin{document}

\title{An improved watermarking scheme for Internet applications}
\author{\IEEEauthorblockN{Christophe Guyeux\IEEEauthorrefmark{1} and
Jacques M. Bahi\IEEEauthorrefmark{1}}
\IEEEauthorblockA{\IEEEauthorrefmark{1}University of Franche-Comté\\
Computer Science Laboratory LIFC,
Besan\c con, France\\ Email: \{christophe.guyeux, jacques.bahi\}@univ-fcomte.fr}}

\maketitle

\begin{abstract}
In this paper, a data hiding scheme ready for Internet applications is proposed. An existing scheme based on chaotic iterations is improved, to respond to some major Internet security concerns, such as digital rights management, communication over hidden channels, and social search engines. By using Reed Solomon error correcting codes and wavelets domain, we show that this data hiding scheme can be improved to solve issues and requirements raised by these Internet fields.
\end{abstract}
\IEEEpeerreviewmaketitle

\section{Introduction}

Information hiding has recently become a major digital technology, especially with the increasing importance and widespread distribution of digital media through the Internet. It encompasses steganography\cite{Ker06} and digital watermarking. The aim of watermarking is to slightly alter some digital documents, like pictures or movies, for a large panel of reasons, such as: copyright protection, control utilization, data description, integrity checking, or content authentication. Various qualities are required for a watermarking method, depending on the aims to reach: discretion, robustness against attacks~\cite{AdelsbachKS06}, \emph{etc.} Many watermarking schemes have been proposed in recent years, which can be classified into two categories: spatial domain \cite{Wu2007} and frequency domain watermarking \cite{Cong2006}, \cite{Dawei2004}. In spatial domain watermarking, a great number of bits can be embedded without inducing too clearly visible artifacts, while frequency domain watermarking has been shown to be quite robust against JPEG compression, filtering, noise pollution and so on. More recently, chaotic methods have been proposed to encrypt the watermark, or embed it into the carrier image, to improve security. 

Information hiding is now an integral part of Internet technologies. In the field of social search engines, for example, contents like pictures or movies are tagged with descriptive labels by contributors, and search results are determined by these descriptions. These collaborative taggings, used for example in Flickr~\cite{Frick} and Delicious~\cite{Delicious} websites, contribute to the development of a Semantic Web, in which every Web page contains machine-readable metadata that describe its content. Information hiding technologies can be used for embedding these metadata. The advantage of its use is the possibility to realize  social search without websites and databases: descriptions are directly embedded into media, whatever their formats. Robustness is required in this situation, as descriptions should resist to modifications like resizing, compression, and format conversion.

The Internet security field is also concerned by watermarking technologies. Steganography and cryptography are supposed to be used by terrorists to communicate through the Internet. Furthermore, in the areas of defense or in industrial espionage, many information leaks using steganographic techniques have been discovered. Lastly, watermarking is often cited as a possible solution to digital rights managements issues, to counteract piracy of digital work in an Internet based entertainment world~\cite{Nakashima2003}.

In this paper, the desire is to improve the robustness of the watermarking scheme proposed in~\cite{guyeux10ter}, to respond to Internet security concerns recalled above. The robustness of the watermarking scheme through geometric attacks is improved by using Reed Solomon correcting codes, whereas the capacity to withstand JPEG compression and noise pollution attacks is enlarged by embedding the watermark into the wavelets domain. Due to its improved robustness, this scheme is suitable for tagging multimedia contents in a social web search context. Additionally, the proposed scheme possesses various properties of chaos and is secure (see~\cite{guyeux10ter}), so it is suitable when desiring to establish a hidden communication channel through the Internet, or for digital rights management. Lastly, watermark encryption and authentication are possible, which enlarge the variety of use in Internet security applications.

The rest of this paper is organized as follows. Firstly, some basic definitions concerning chaotic iterations and topological chaos are given in Section~\ref{Section:Basic recalls}. The data hiding scheme used in this paper is recalled in the same section. In Section~\ref{Geometric}, the way to use Reed Solomon error correcting codes to improve robustness against geometric attacks is given. Then it is explained in Section~\ref{frequency} how to improve robustness against frequency domain attacks by using wavelets coefficients into our scheme. The paper ends with a conclusion section where the contribution is summed up and the planned future work is discussed.

\section{Basic recalls}
\label{Section:Basic recalls}

This section is devoted to the recall of the data hiding scheme, which will be improved in Sections~\ref{Geometric} and~\ref{frequency}. To do so, basic notations and terminologies in the fields of chaotic iterations and topological chaos are introduced.

\subsection{Chaotic iterations and Devaney's chaos}

\subsubsection{Chaotic iterations}

In the sequel $S^{n}$ denotes the $n^{th}$ term of a sequence $S$, $V_{i}$ denotes the $i^{th}$ component of a vector $V$, and $f^{k}=f\circ ...\circ f$ is for the $k^{th}$ composition of a function $f$. Finally, the following notation is used: $\llbracket1;N\rrbracket=\{1,2,\hdots,N\}$.

Let us consider a \emph{system} of a finite number $\mathsf{M}$ of \emph{cells}, so that each cell has a boolean \emph{state}. Then a sequence of length $\mathsf{M}$ of boolean states of the cells corresponds to a particular \emph{state of the system}. A sequence which elements belong in $\llbracket 1;\mathsf{M} \rrbracket $ is called a \emph{strategy}. The set of all strategies is denoted by $\mathbb{S}.$

\begin{definition}
Let $S\in \mathbb{S}$. The \emph{shift} function is defined by $\sigma :(S^{n})_{n\in \mathds{N}}\in \mathbb{S}\longrightarrow (S^{n+1})_{n\in \mathds{N}}\in \mathbb{S}$ and the \emph{initial function} $i$ is the map which associates to a sequence, its first term: $i:(S^{n})_{n\in \mathds{N}}\in \mathbb{S}\longrightarrow S^{0}\in \llbracket1;\mathsf{M}\rrbracket$.
\end{definition}

\begin{definition}

The set $\mathds{B}$ denoting $\{0,1\}$, let $f:\mathds{B}^{\mathsf{M}%
}\longrightarrow \mathds{B}^{\mathsf{M}}$ be a function and $S\in \mathbb{S}
$ be a strategy. Then, the so-called \emph{chaotic iterations} are defined~\cite{Robert1986} by $x^0\in \mathds{B}^{\mathsf{M}}$ and $\forall n\in \mathds{N}^{\ast }$, $\forall i\in \llbracket1;\mathsf{M}\rrbracket$,

\begin{equation}
\begin{array}{l}
x_i^n=\left\{
\begin{array}{ll}
x_i^{n-1} & \text{ if }S^n\neq i, \\
\left(f(x^{n-1})\right)_{S^n} & \text{ if }S^n=i.%
\end{array}%
\right.%
\end{array}%
\end{equation}
\end{definition}

In other words, at the $n^{th}$ iteration, only the $S^{n}-$th cell is \textquotedblleft iterated\textquotedblright .

\subsubsection{Devaney's chaotic dynamical systems}
\label{Devaney's chaotic dynamical systems}

Consider a metric space $(\mathcal{X},d)$ and a continuous function $f:%
\mathcal{X}\longrightarrow \mathcal{X}$. $f$ is said to be \emph{%
topologically transitive} if, for any pair of open sets $U,V\subset \mathcal{%
X}$, there exists $k>0$ such that $f^{k}(U)\cap V\neq \varnothing $. $(%
\mathcal{X},f)$ is said to be \emph{regular} if the set of periodic points
is dense in $\mathcal{X}$. $f$ has \emph{sensitive dependence on initial
conditions} if there exists $\delta >0$ such that, for any $x\in \mathcal{X}$
and any neighborhood $V$ of $x$, there exists $y\in V$ and $n\geqslant 0$
such that $|f^{n}(x)-f^{n}(y)|>\delta $. $\delta $ is called the \emph{%
constant of sensitivity} of $f$.

Quoting Devaney in \cite{Devaney}, a function $f:\mathcal{X}\longrightarrow \mathcal{X}$ is said to be ``chaotic'' on $\mathcal{X}$ if $(\mathcal{X} ,f)$ is regular, topologically transitive, and has sensitive dependence on initial conditions.
When $f$ is chaotic, then the system $(\mathcal{X}, f)$ is highly unpredictable because of regularity and sensitive dependence on initial conditions. Moreover, it cannot be simplified (broken down or decomposed into two subsystems which do not interact) because of topological transitivity. These chaotic dynamical systems then present behaviors very similar to physical noise sources.

In~\cite{guyeux10}, a rigorous theoretical framework has been introduced for the study of chaotic iterations. It has been proven that chaotic iterations (CIs) presented above satisfy topological chaos properties, which leads to improve the security of data hiding schemes based on CIs.

\subsection{Definition of a chaos-based data hiding scheme}
\label{sec:Algo}

\subsubsection{Most and least significant coefficients}

Let us define the notions of most and least significant coefficients of an image.

\begin{Definition}
\label{definitionMSC}
For a given image, most significant coefficients (in short MSCs), are coefficients that allow the description of the relevant part of the image, \emph{i.e.}, its richest part (in terms of embedding information), through a sequence of bits.
\end{Definition}

\begin{figure}[htb]

\begin{minipage}[b]{1.0\linewidth}
  \centering
 \centerline{\includegraphics[width=3.5cm]{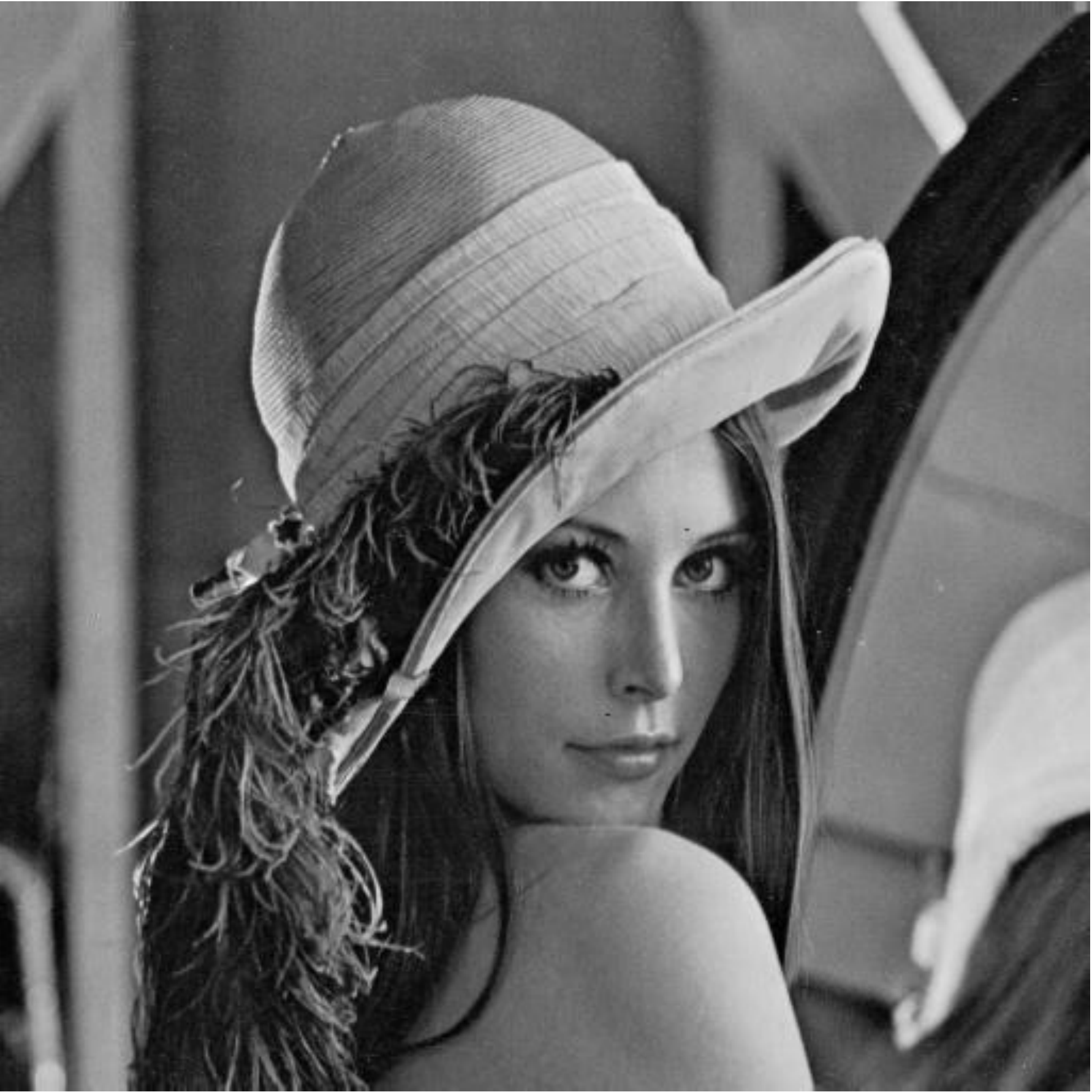}}
  \centerline{(a) Lena.}
\end{minipage}

\begin{minipage}[b]{.48\linewidth}
  \centering
  \centerline{\includegraphics[width=3.5cm]{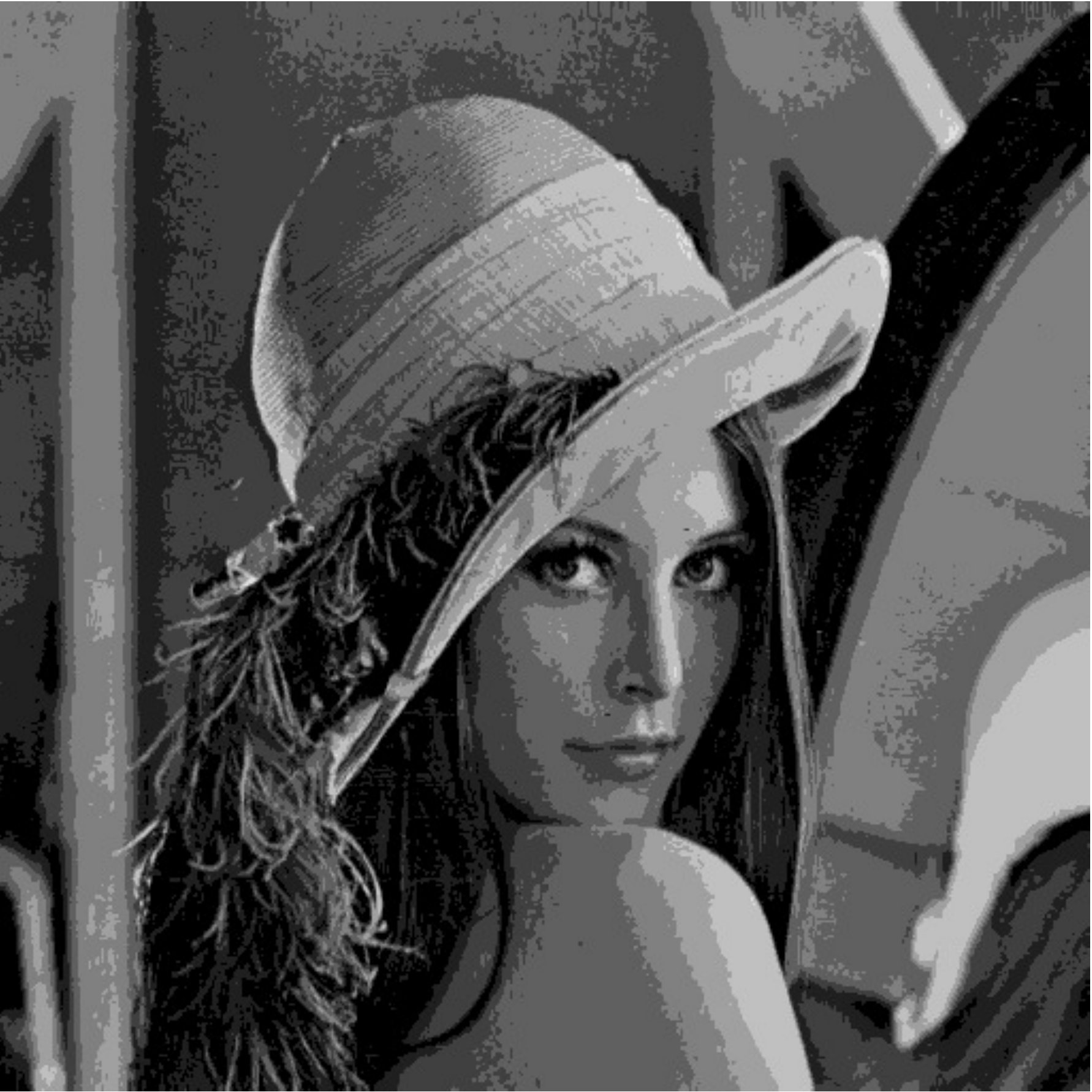}}
  \centerline{(b) MSCs of Lena.}
\end{minipage}
\hfill
\begin{minipage}[b]{0.48\linewidth}
  \centering
 \centerline{\includegraphics[width=3.5cm]{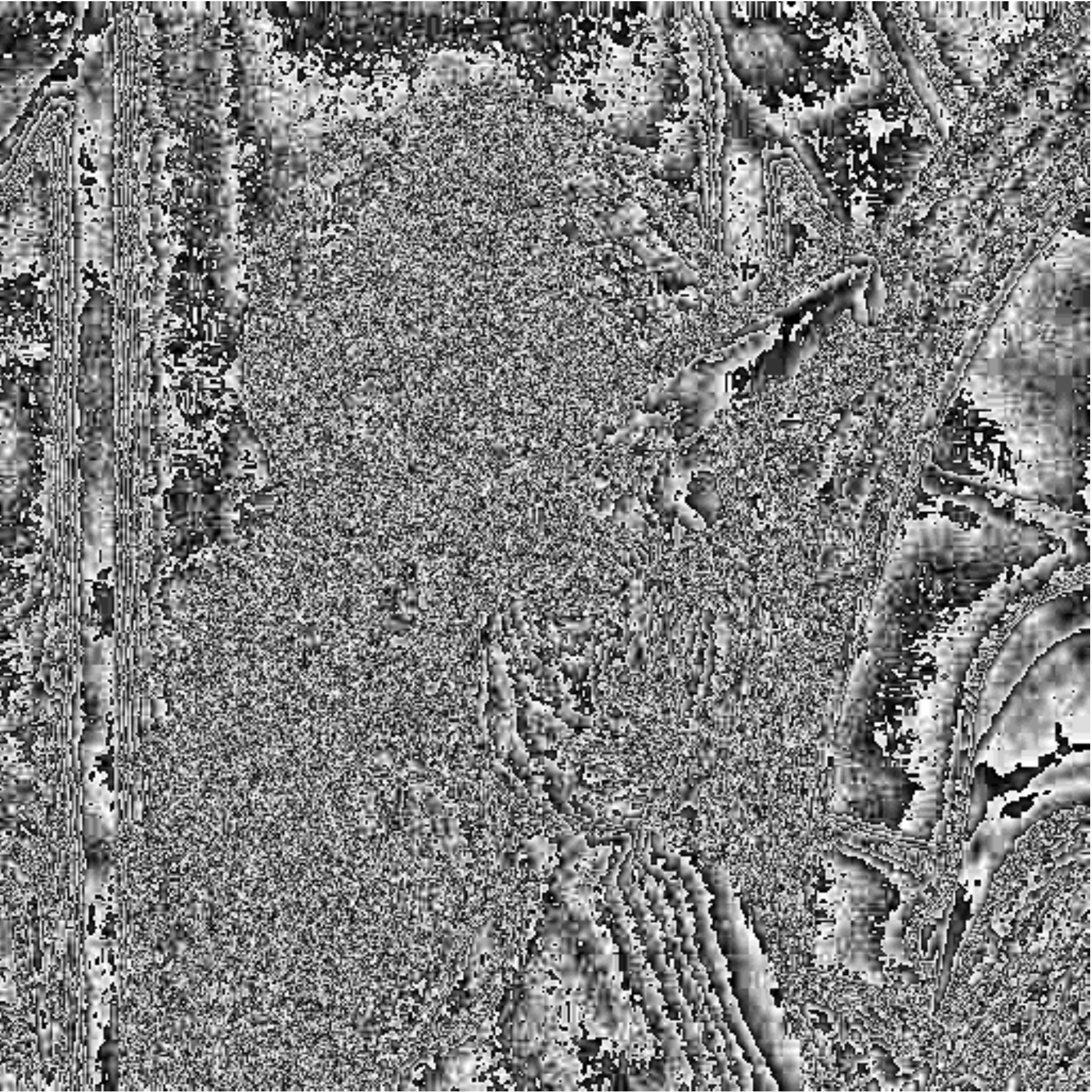}}
  \centerline{(c) LSCs of Lena ($\times 17$).}
\end{minipage}
\caption{Example of most and least significant coefficients of Lena.}
\label{fig:MSCLC}
\end{figure}

For example, in a spatial description of a grayscale image, a definition of MSCs can be the sequence constituted by the first four bits of each pixel (see Figure~\ref{fig:MSCLC}). In a discrete cosine frequency domain description, each $8\times 8$ block of the carrier image is mapped onto a list of 64 coefficients. The energy of the image is mostly contained in a determined part of themselves, which can constitute a possible sequence of MSCs.

\begin{Definition}
\label{definitionLSC}
By least significant coefficients (LSCs), we mean a translation of some insignificant parts of a medium in a sequence of bits (insignificant can be understand as: ``which can be altered without sensitive damages'').
\end{Definition}

These LSCs can be, for example, the last three bits of the gray level of each pixel (see Figure~\ref{fig:MSCLC}). Discrete cosine, Fourier, and wavelet transforms can be used also to generate LSCs and MSCs. Moreover, these definitions can be extended to other types of media.


LSCs are used during the embedding stage. Indeed, some of the least significant coefficients of the carrier image will be chaotically chosen and switched, or replaced by the bits of the watermark. The MSCs are only useful in case of authentication; mixture and embedding stages depend on them. Hence, a coefficient should not be defined at the same time as a MSC and a LSC: the last can be altered while the first is needed to extract the watermark.

\subsubsection{Stages of the scheme}

Our data hiding scheme consists of two stages: (1) mixture of the watermark and (2) its embedding.

\paragraph{Watermark mixture}

Firstly, for safety reasons, the watermark can be mixed before its embedding into the image. A common way to achieve this stage is to use the bitwise exclusive or (XOR), for example between the watermark and a pseudo-random binary sequence provided by the generator defined in~\cite{guyeux09bis}. In this paper, we introduce a new mixture scheme based on chaotic iterations. Its chaotic strategy will be highly sensitive to the MSCs, in the case of an authenticated watermarking.

\paragraph{Watermark embedding}

Some LSCs will be switched, or substituted by the bits of the possibly mixed watermark. To choose the sequence of LSCs to be altered, a number of integers, less than or equal to the number $\mathsf{M}$ of LSCs corresponding to a chaotic sequence $U$, is generated from the chaotic strategy used in the mixture stage. Thus, the $U^{k}$-th least significant coefficient of the carrier image is either switched, or substituted by the $k^{th}$ bit of the possibly mixed watermark. In case of authentication, such a procedure leads to a choice of the LSCs which are highly dependent on the MSCs~\cite{guyeux10}.

On the one hand, when the switch is chosen, the watermarked image is obtained from the original image whose LSBs $L = \mathds{B}^{\mathsf{M}}$ are replaced by the result of some chaotic iterations. Here, the iterate function is the vectorial boolean negation,
\begin{equation}
f_0:(x_1,...,x_\mathsf{M}) \in \mathds{B}^\mathsf{M} \longmapsto (\overline{x_1},...,\overline{x_\mathsf{M}}) \in \mathds{B}^\mathsf{M},
\end{equation}
the initial state is $L$, and the strategy is equal to $U$. In this case, the whole embedding stage satisfies topological chaos properties (see~\cite{guyeux10}), but the original medium is needed to extract the watermark. On the other hand, when the selected LSCs are substituted by the watermark, its extraction can be done without the original cover (blind watermarking). In this case, the selection of LSBs still remains chaotic because of the use of a chaotic map, but the whole process does not satisfy topological chaos~\cite{guyeux10}. The use of chaotic iterations is reduced to the mixture of the watermark. See the following sections for more detail.

\paragraph{Extraction}

The chaotic strategy can be regenerated even in the case of an authenticated watermarking, because the MSCs have not been changed during the embedding stage. Thus, the few altered LSCs can be found, the mixed watermark can be rebuilt, and the original watermark can be obtained. In case of a switch, the result of the previous chaotic iterations on the watermarked image should be the original cover. The probability of being watermarked decreases when the number of differences increase.

If the watermarked image is attacked, then the MSCs will change. Consequently, in case of authentication and due to the high sensitivity of the embedding sequence, the LSCs designed to receive the watermark will be completely different. Hence, the result of the recovery will have no similarity with the original watermark.

\begin{figure}[htb]
\centerline{\includegraphics[width=8cm]{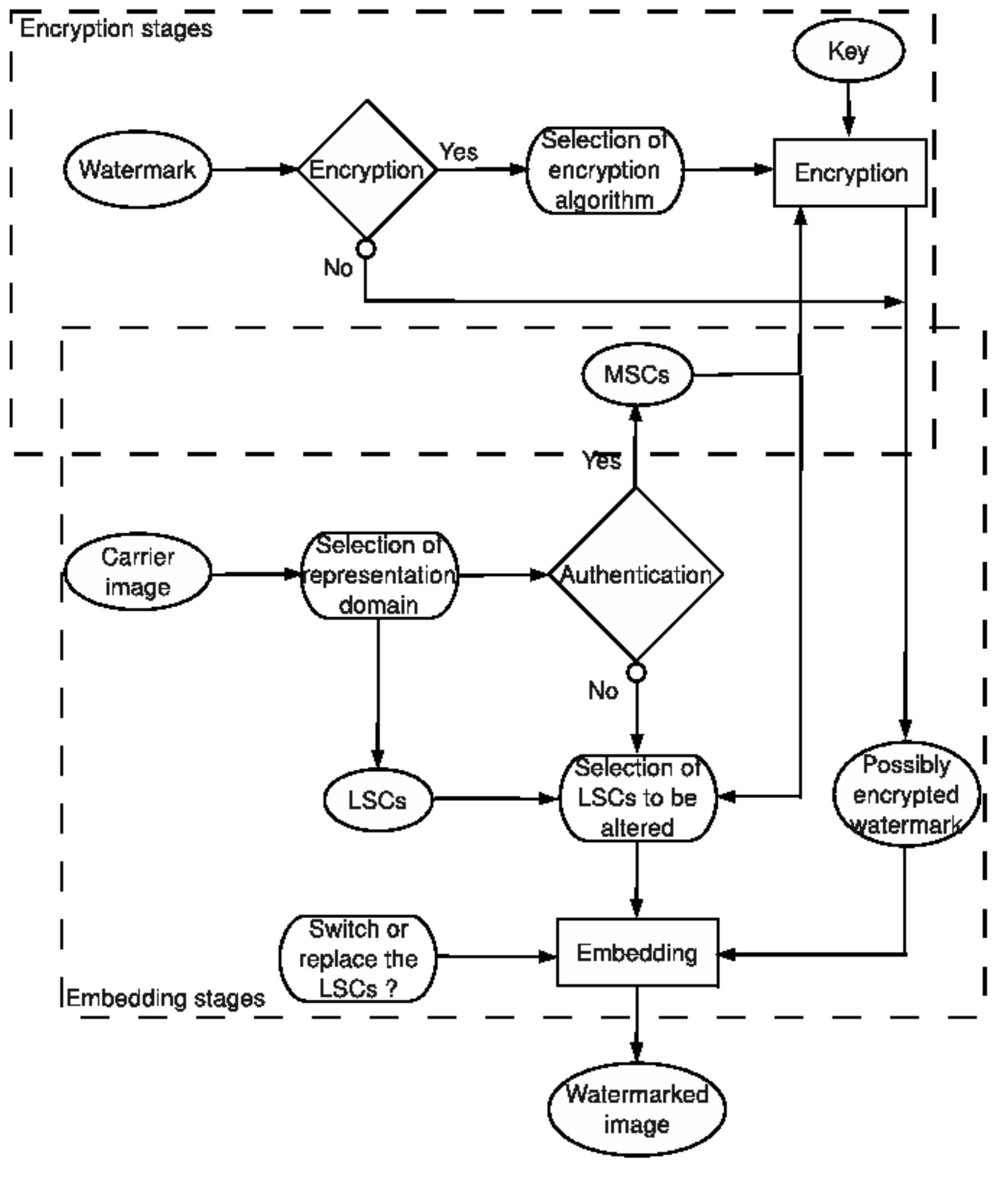}}
\caption{The chaos-based data hiding decision tree.}
\label{fig:organigramme}
\end{figure}

The chaos-based data hiding scheme is summed up in Figure~\ref{fig:organigramme}.

\section{Improving robustness against geometric attacks}
\label{Geometric}

In this section, we are interested in improving our scheme to make its use relevant in a social web search context. The idea is to embed the tag of a given image into its pixel values. As neither the cover image nor the tag should be required during a search, the LSBs will be replaced (not switched), by the tags. Authentication is not required, as man-in-the-middle attacks are not supposed to occur. However tags are vulnerable to involuntary attacks such as rotation or resizing, so we need to improve robustness against geometric attacks. To do so, the embedding domain will be the spatial domain. This choice leads to a large given payload, so a lot of tags can be embedded into the cover image. Additionally, we will use Reed-Solomon error correcting codes, to reinforce the capacity to extract the watermark from a tagged image, even though it has been altered.
As an illustrative example, we will show how to embed the description of the well-known Lena into its own image. 
Let us remark that the same procedure can be applied to create a hidden channel for communicating through a given web page, for example, by inserting messages in the background image of this website. In this situation, it is recommended to add the encryption stage to improve the security of the hidden channel.

In this illustrative example, the following text extracted from Wikipedia's description of Lena will be inserted into its own image:

\begin{alltt}
Lena (Soderberg), a standard test image
originally cropped from the November 
1972 issue of Playboy magazine.
\end{alltt}

The cover image will be the Figure~\ref{fig:MSCLC}(a), which is a $256 \times 256$ grayscale image. The text to embed is converted into 756 binary digits by using the ASCII table: each of the 109 characters are coded with 7 digits, thus obtaining the following bits flow (called a system):

\begin{small}
\noindent
100110011001011101110110000101000000101000101001111011
111100100110010111100101100010110010111100101100111...
\end{small}

20000 binary digits are computed from a logistic map, with parameters $\mu = 3.999999$, $x^0 = 0.65$, and those binary digits are grouped ten by ten ($10 = \lceil \log_2(756)\rceil$) to obtain an integer sequence $S$ lesser than or equal to 756. So, chaotic iterations are applied to the above system, with chaotic strategy $S$ and the vectorial boolean negation, to obtain the following encrypted message:

\begin{small}
\noindent
001000111110001110001101110111111000011011010011000101
001011110000110110011010010001110101101100010110101...
\end{small}

In this example, there is no authentication step, but Reed-Solomon error correction codes are used to increase the robustness. Here, two layers of Reed–Solomon coding, respectively (32,24)-RS and (24,16)-RS codes, are separated by a 3-way convolutional interleaver operation, to obtain a scheme similar to the Cross-Interleaved Reed Solomon Coding (CIRC) of the compact disc. The message to embed is the result of this coding operation: a 2112 binary stream, starting by:

\begin{small}
\noindent
010110100101100000100001000111000010011100111111010001
110111100000010110001101010111011000010011001001110...
\end{small}

These 2112 bits will be embedded into Lena, an image constituted by $256 \times 256 \times 8 = 524288$ bits (8 bits per pixel). To do so, we will consider the two least significant bits of each pixel as LSCs: a few of them will be replaced by the bits of the watermark. To select these bits to replace, the strategy $S$ of the encryption stage is used again, to generate a sequence of triplets $(x^n, y^n, z^n)_{n\in \mathds{N}}$ in such a way that $x^n, y^n \in \llbracket 0 ; 255 \rrbracket^\mathds{N}$, and $z^n \in \{1 ; 2 \}^\mathds{N}$. This generation is realized as follows:

$$
\left\{
\begin{array}{lll}
x^0 & = & 11, \\
y^0 & = & 23, \\
z^0 & = & 1, \\
\end{array}
\right.
$$

and

$$
\left\{
\begin{array}{llll}
x^{n+1} & = & 2 x^n +S^{3n}+n & (\text{mod } 255), \\
y^{n+1} & = & 2 y^n +S^{3n+1}+n & (\text{mod } 255), \\
z^{n+1} & = & 2 z^n +S^{3n+2}+n & (\text{mod } 2). \\
\end{array}
\right.
$$

\begin{figure}[htb]

\begin{minipage}[b]{1.0\linewidth}
  \centering
 \centerline{\includegraphics[width=3.5cm]{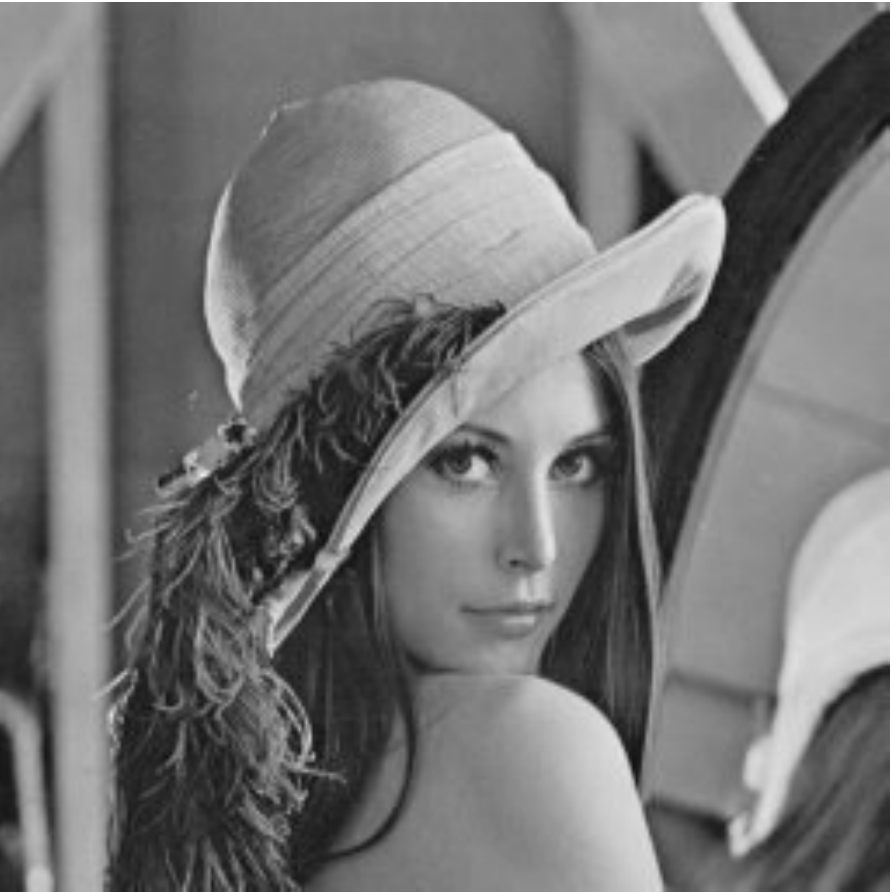}}
  \centerline{(a) Watermarked Lena.}
\end{minipage}
\begin{minipage}[b]{.48\linewidth}
  \centering
 \centerline{\includegraphics[width=3.5cm]{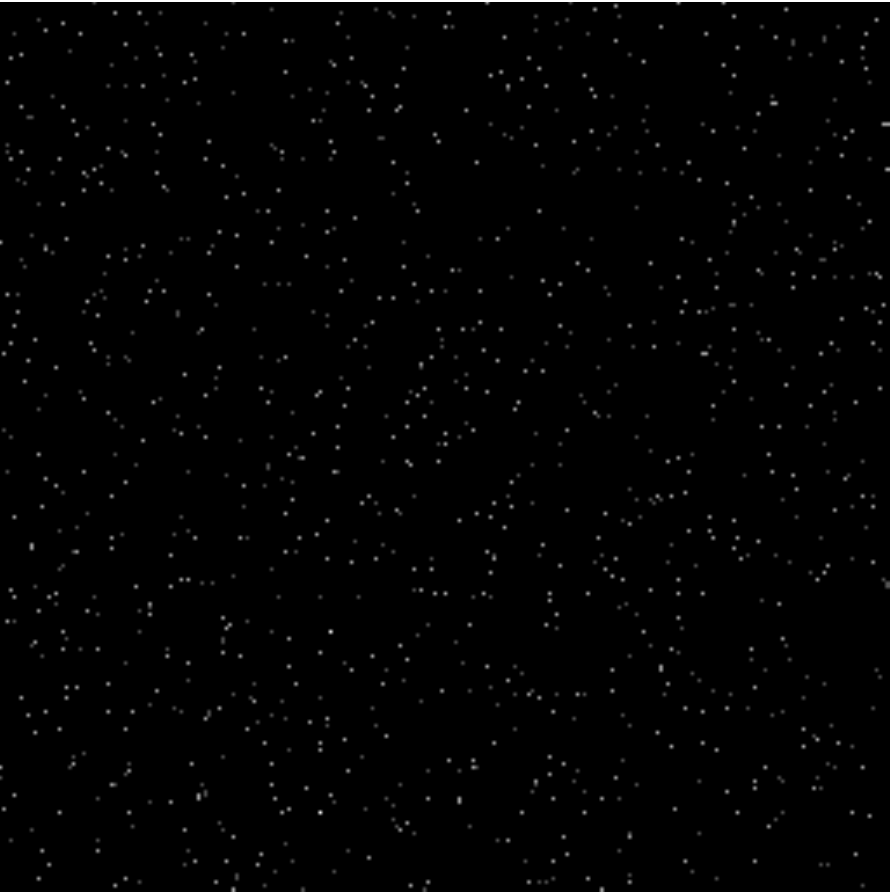}}
  \centerline{(b) Differences with Lena.}
\end{minipage}
\hfill
\begin{minipage}[b]{0.48\linewidth}
  \centering
 \centerline{\includegraphics[width=3.5cm]{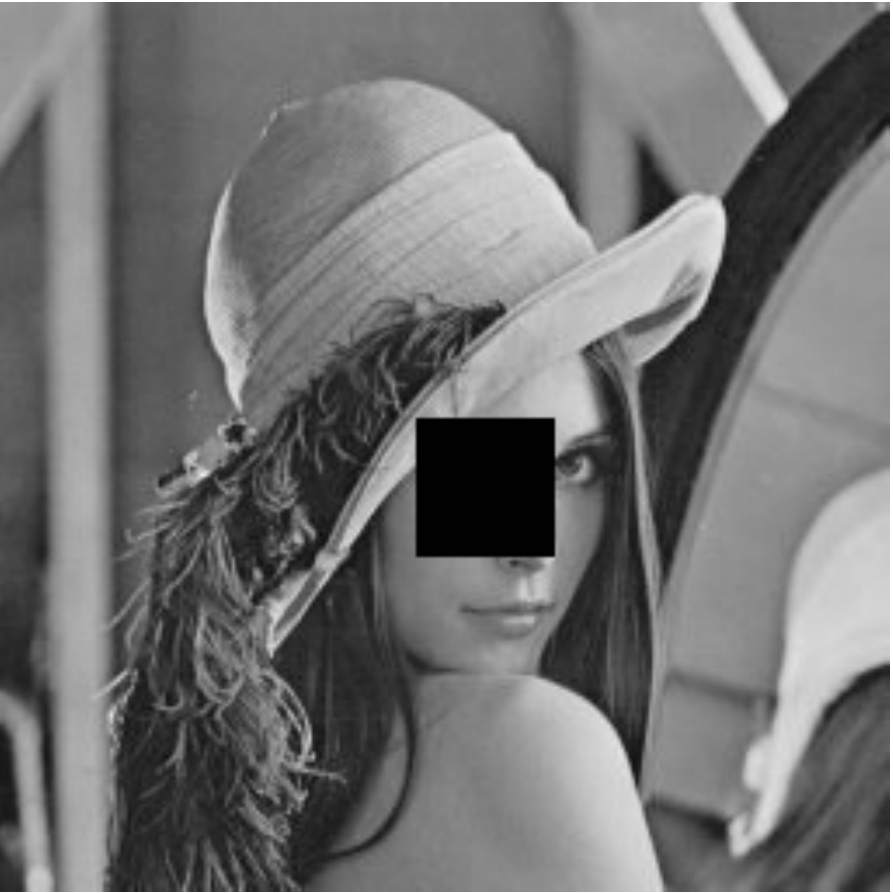}}
  \centerline{(c) Attacked Lena.}
\end{minipage}
\caption{Watermarked Lenas (scale reduced).}
\label{fig:water}
\end{figure}

So the $n^{th}$ bit of the encrypted and encoded binary message is inserted into the $z^n$ least significant bit of the pixel in position $(x^n, y^n)$ of Lena, to obtain the watermarked Lena in Figure~\ref{fig:water}(a).
In Figure~\ref{fig:water}(b) the differences are shown between the original Lena and the watermarked Lena.
This image illustrates the fact that LSCs to be replaced are chaotically chosen and uniformly distributed~\cite{guyeux10}.

In~\cite{guyeux10ter}, our scheme has been defined without RS codes and its robustness has been evaluated. It is established that the watermark can resist rotation, cropping, JPEG compression, and gaussian noise attacks. However, the extracted watermark is slightly different from the original one and this difference increases with the number of attacks. These errors, which are undesirable in a social web search context, can be corrected by the use of RS codes. 

To illustrate, the watermarked Lena is zeroed: a square of $40\times 40$ pixels is removed from the image, as in Figure~\ref{fig:water}(c). So the message is extracted from the watermarked and attacked Lena: the strategy $S$ is regenerated from a logistic map with the same parameters as above. Then the sequences $x^n, y^n$ and $z^n$ can be regenerated too, and the embedded bits can thus be extracted. These bits are decoded in the reverse process: (24,16)-RS decoding, 3-way de-interlacing, and (32,24)-RS decoding codes.
Lastly, the resulting bits sequence is decrypted, bits are grouped 7 by 7, and converted into characters with the ASCII table, to obtain the following message:

\begin{alltt}
Lena (Soderberg), a standard test image
originally cropped from the November 
1972 issue of Playboy magazine.
\end{alltt}

\section{Improving robustness against frequency domain attacks}
\label{frequency}

In this section, the way to use our scheme in frequency DWT domain is explained. Due to its robustness against frequency attacks such as JPEG compression, this scheme can be used to insert a copyright into a media (digital rights management context).

\subsection{Stages and detail}

The carrier image and watermark are the same as in Section~\ref{Geometric}, but Lena is now constituted by $512 \times 512$ pixels. The embedding domain is the discrete wavelets domain (DWT). In this paper, the Daubechies family of wavelets is chosen: Lena is converted into its Daubechies-1 DWT coefficients, which are altered by chaotic iterations. The watermark is encrypted by chaotic iterations before its embedding, with the same procedure as above.

\begin{figure}[htb]
\centerline{\includegraphics[width=3.7cm]{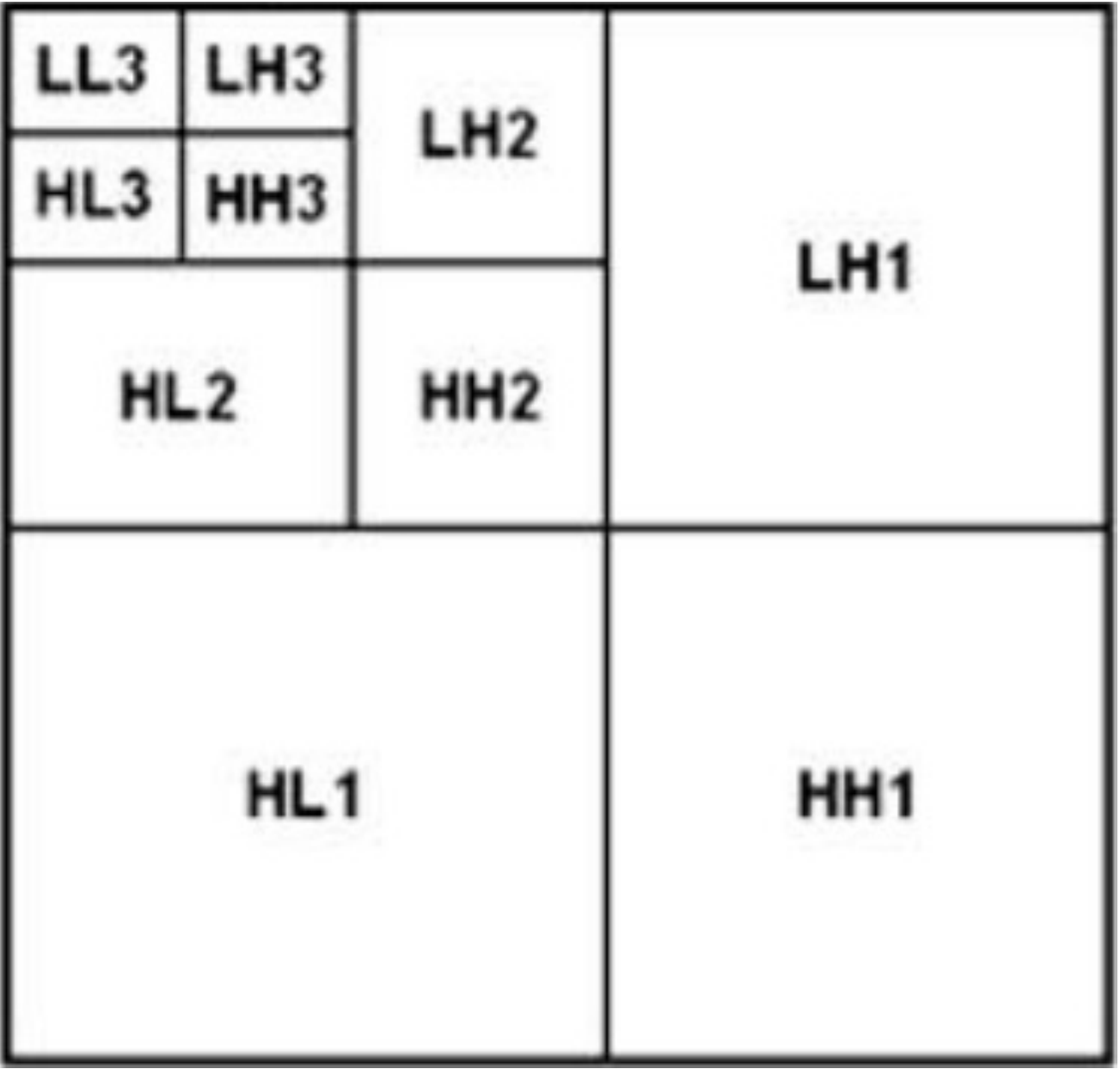}}
\caption{Wavelets coefficients.}
\label{fig:DWTs}
\end{figure}

Each example below depends on a decomposition level and a coefficient matrix (Figure~\ref{fig:DWTs}): $LL$ means approximation coefficient, when $HH,LH,HL$ denote respectively diagonal, vertical and horizontal detail coefficients. For example, the DWT coefficient HH2 is the matrix equal to the diagonal detail coefficient of the second level of decomposition of Lena.

To embed the encrypted watermark, LSCs are obtained from the coefficients defined above. The system to iterate is the boolean vector of size $256^2$, constituted by these \textsf{M} LSCs of Lena. Iterate function is the vectorial boolean negation, and chaotic strategy $U$ is defined as follows:
\begin{equation}
\left\{ 
\begin{array}{lll}
U^{0} & = & S^{0} \\ 
U^{n+1} & = & S^{n+1}+2\times U^{n}+n ~ (\textrm{mod } \textsf{M}).
\end{array}%
\right.
\label{equation_strategie}
\end{equation}

\noindent where $S$ denotes the strategy used in the encryption stage (see~\cite{guyeux10ter}). Thus, bits of the LSCs are switched, not replaced: the whole embedding process satisfies Devaney's chaos property~\cite{guyeux10}. However, for this reason, the watermark cannot be extracted: contrary to Section~\ref{Geometric}, we are not in a steganographic framework, but in a pure non-blind watermarking scheme used for digital rights management. To know if a given image $I'$ is the watermarked version of another image $I$:
\begin{itemize}
\item the whole process is applied to $I'$, with the same parameters (LSCs, watermark, \emph{etc.}), to obtain $I''$,
\item $I''$ is compared to the original $I$.
\end{itemize}

To evaluate the differences, the RMS value defined by $\bar{x} = \sqrt{\frac{1}{\mathsf{M}}\sum_{i=1}^\mathsf{M}{(I-I'')_i^2}}$ is computed. The probability that the image has been watermarked increases when the RMS decreases. Indeed, each bit of the LSCs of $I''$ has been switched an even number of times (the RMS is nonzero because of computational errors).

\subsection{First example: coefficient HH2}

\subsubsection{Embedding}

In this first experiment, the watermark is inserted into the diagonal coefficient HH2 (a real matrix of size $128 \times 128$). LSCs are the second least significant bit of each integral value of HH2. To do the insertion, chaotic iterations are made. The system to iterate is the boolean vector of size $128^2$, constituted by the LSCs of Lena. Iterate function is the vectorial boolean negation and chaotic strategy is defined as in Equation~\ref{equation_strategie}, with $U^0 = 1$ and $\mathsf{M} = 256^2$.

\begin{figure}[htb]
\begin{minipage}[b]{.45\linewidth}
  \centering
 \centerline{\includegraphics[width=3.5cm]{lena512}}
  \centerline{(a) Original Lena.}
\end{minipage}
\hfill
\begin{minipage}[b]{0.45\linewidth}
  \centering
 \centerline{\includegraphics[width=3.5cm]{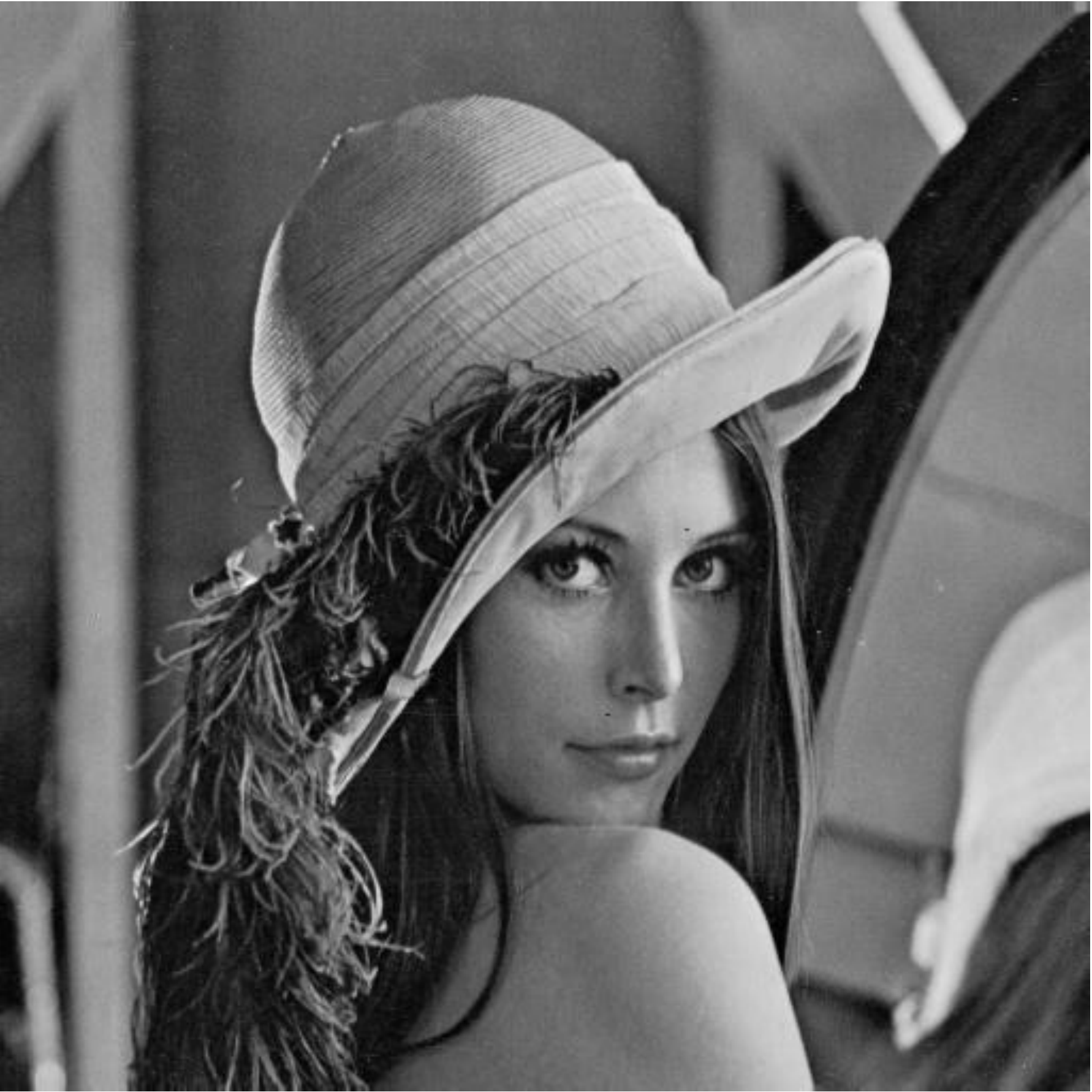}}
  \centerline{(b) Watermarked Lena.}
\end{minipage}
\caption{Data hiding in DWT domain}
\label{fig:DWT}
\end{figure}

In this situation, PSNR = 53.45 dB. Pixel values have been modified by at most of one level of gray. The mean value of differences is equal to 0.294, when RMS = 0.542. The alteration can thus be considered as indistinguishable.

\subsubsection{Extraction}

The system to iterate is constituted by the second least significant bit of each integral value of HH2, the approximation coefficient of the first decomposition level of the watermarked Lena. The iterate function is the vectorial boolean negation and the chaotic strategy is computed as above. Thus, the result is compared to the coefficient HH2 of the original Lena. The RMS is equal to 0.129.
As a comparison, Table~\ref{Tab:RMS} gives the RMS values resulting on a bad extraction (wrong parameters, \emph{etc.}) Symbol `-' means that the value of the considered parameter is unchanged.
We show that the least RMS is obtained for an extraction with the same parameters as the embedding. Let us notice that if the extraction is attempted to the original Lena, RMS is twice greater than 0.127.

\begin{table}
\renewcommand{\arraystretch}{1.3}
\centering
\caption{RMS values for a HH2 embedding}
\label{Tab:RMS}

{\scriptsize

\begin{tabular}{c c c c c c}

\toprule

& \multicolumn{5}{c}{HH2 embedding} \\ \midrule

& $\mu$ & $U^0$ & Iterations & Authentication & RMS \\ \cmidrule(r){2-6}

& 3.99987 & - & - & - & 1.131 \\

& - & 0.64 & - & - & 1.129 \\

Encryption & - & - & 19950 & - & 0.796 \\

 & - & - & - & MSB = [5,6,7] & 1.122 \\ \midrule

& Coefficient & $S^0$ & LSB & RMS \\ \cmidrule(r){2-5}

& HH1 & - & - & 253.65 \\

Embedding & - & 2 & - & 0.653 \\

& - & - & [1] & 0.983 \\ \bottomrule

\end{tabular}} 
\end{table}

\subsection{Second example: coefficient LL1}

\subsubsection{Embedding}

In this paragraph, the watermark is inserted into the approximation coefficient LL1 of Lena (a real matrix of size $256 \times 256$) and LSCs are the second least significant bit of each integral value of LL1.

To realize the embedding, chaotic iterations are realized as before. The system to iterate is the boolean vector, of size $256^2$, constituted by the LSCs of Lena. Iterate function is the vectorial boolean negation, chaotic strategy is defined as in Equation~\ref{equation_strategie} with $U^0 = 1$, and $\mathsf{M} = 256^2$.
In this situation, PSNR = 60.06 dB. Pixel values have been modified by at most two levels of gray. The mean value of differences is 0.063, when the RMS is equal to 0.245. For all of these reasons, the alteration can be considered again as indistinguishable.

\subsubsection{Extraction}

The system to iterate is constituted by the second least significant bit of each integral value of LL1. Iterate function is the vectorial boolean negation and chaotic strategy is computed as above. Thus, the result is compared to the coefficient LL1 of the original Lena. In our example, we obtain RMS = 0.127.
As a comparison, Table~\ref{Tab:RMS2} gives the RMS values resulting in a bad extraction (wrong parameters, \emph{etc.}) Symbol `-' means that the value of the considered parameter is unchanged.
We show that the least RMS is obtained for an extraction with the same parameters as the embedding. Let us remark that if the extraction is tried on the original Lena, then RMS is twice greater than $0.127$.

\begin{table}
\renewcommand{\arraystretch}{1.3}
\centering
\caption{RMS values for a LL1 embedding}
\label{Tab:RMS2}

{\scriptsize

\begin{tabular}{c c c c c c}

\toprule

& \multicolumn{5}{c}{LL1 embedding} \\ \midrule

& $\mu$ & $U^0$ & Iterations & Authentication & RMS \\ \cmidrule(r){2-6}

& 3.99987 & - & - & - & 0.669 \\

& - & 0.64 & - & - & 0.670 \\

Encryption & - & - & 19950 & - & 0.443 \\

 & - & - & - & MSB = [5,6,7] & 0.667 \\ \midrule

& Coefficient & $S^0$ & LSB & RMS \\ \cmidrule(r){2-5}

& HH1 & - & - & 223.737 \\

Embedding & - & 2 & - & 0.135 \\

& - & - & [1] & 0.548 \\ \bottomrule

\end{tabular}} 
\end{table}

\section{Discussion and future work}

In this paper, the robustness of the data hiding scheme proposed in~\cite{guyeux10ter} is improved to achieve properties required in Internet applications of data hiding techniques. This scheme depends on a general description of the carrier medium to watermark, in terms of the significance of some coefficients we called MSCs and LSCs. The encryption of the watermark and the selection of coefficients to alter are based on chaotic iterations, which generate topological chaos in the sense of Devaney~\cite{guyeux10}. Thus, the proposed scheme has a sufficient level of security for Internet applications, such as digital rights management or social web search.

We have proposed in this paper to enlarge the relevance of our scheme in these contexts by using Reed-Solomon error correcting codes and wavelets domain. The first improvement is relevant in a social web search domain, in which the tags of an image must be recovered exactly, even though the image has faced geometric operations. The use of wavelets domain is linked more to digital rights management. This domain is known to present good results against frequency attacks, which can occur when someone tries to remove some DRM. It can be noticed that these two improvements can be realized together.

The schemes have been evaluated through attacks and results have been experimentally obtained. Choices that have been made in this first study are simple: spatial and Daubechies domains for the embedding, negation function as iteration function, \emph{etc.} The aim was not to find the best watermarking method generated by our general scheme, but to explain how to improve robustness for Internet applications.

In future work, other choices of iteration functions and chaotic strategies will be explored and compared in order to increase authentication and robustness to attacks. In addition, new frequency domain representations will be used to select the MSCs and LSCs. Properties induced by topological chaos, such as entropy, will be explored and their role in Internet applications will be explained.

\bibliographystyle{plain}
\bibliography{internet2010papier2}

\end{document}